\documentclass[conference]{IEEEtran}

\newtheorem{thm}{Theorem}
\newtheorem{lem}[thm]{Lemma}
\newtheorem{cor}[thm]{Corollary}
\newtheorem{pro}[thm]{Proposition}

\usepackage{cite}

\ifCLASSINFOpdf
  \usepackage[pdftex]{graphicx}

\else

   \usepackage[dvips]{graphicx}

\fi

\usepackage{amsmath}

\usepackage{algorithmic}

\usepackage{array}

\usepackage{mathrsfs}
\usepackage{bm}

\ifCLASSOPTIONcompsoc
  \usepackage[caption=false,font=normalsize,labelfont=sf,textfont=sf]{subfig}
\else
  \usepackage[caption=false,font=footnotesize]{subfig}
\fi

\begin{document}
\title{Analysis of Random Pulse Repetition Interval Radar}

\author{\IEEEauthorblockN{Jieli Zhu, Tong Zhao, Tianyao Huang and Dengfeng Zhang}
\IEEEauthorblockA{\IEEEauthorrefmark{1}Radar Research (Beijing), Leihua, Electronic Technology Institute, AVIC, China\\
\IEEEauthorrefmark{2}Aviation Key Laboratory of Science and Technology on AISSS\\
Telephone: (86-10) 84936622, Fax: (86-10) 84936656\\
Email: \{zhujl,zhaotong,huangty,zhangdf\}@ravic.cn}
}

% use for special paper notices
%\IEEEspecialpapernotice{(Invited Paper)}

% make the title area
\maketitle
%\IEEEdisplaynontitleabstractindextext

% As a general rule, do not put math, special symbols or citations
% in the abstract
\begin{abstract}

Random pulse repetition interval (PRI) waveform arouses great interests in the field of modern radars due to its ability to alleviate range and Doppler ambiguities as well as enhance electronic counter-countermeasures (ECCM) capabilities.
Theoretical results pertaining to the statistical characteristics of ambiguity function (AF) are derived in this work, indicating that the range and Doppler ambiguities can be effectively suppressed by increasing the number of pulses and the range of PRI jitters. This provides an important guidance in terms of waveform design.
As is well known, the significantly lifted sidelobe pedestal induced by PRI randomization will degrade the performance of weak target detection. Proceeding from that, we propose to employ orthogonal matching pursuit (OMP) to overcome this issue. Simulation results demonstrate that the OMP method can effectively lower the sidelobe pedestal of strong target and improve the performance of weak target estimation.

\end{abstract}

% no keywords
%\begin{IEEEkeywords}
%Random pulse repetition interval, range ambiguity, doppler ambiguity, target detection.
%\end{IEEEkeywords}

\IEEEpeerreviewmaketitle

\section{Introduction}
\label{sec:intro}
% no \IEEEPARstart
The stable pulse repetition interval (PRI) is commonly used in pulse-Doppler radars for Doppler resolution improvement \cite{Rasool2010}.
The uniformly spaced pulses are processed efficiently through the fast Fourier transform method.
However, due to the periodic characteristics of the pulses, the stable PRI waveforms introduce range and Doppler ambiguities \cite{Levanon2004} and is regarded to have poor electronic counter-countermeasures (ECCM) capabilities \cite{Skolnik2008}.

%Staggered PRI, PRI set and random PRI are three types of stable PRI-based pulse intervals modulation.
Staggered PRI, PRI set and random PRI are three types of pulse intervals modulations evolved from stable PRI.
Staggered PRI contributes to improving the Doppler unambiguous coverage \cite{Thomas1976,Vergara-Dominguez1993}, and the PRI set can solve ambiguity through certain methods such as the Chinese Remainder Theorem \cite{Hovanessian1982,Soyeon2011}.
However, due to the repetitive character of the pulse intervals that still exists \cite{Vergara-Dominguez1993}, these two methods do not lead to a moderate enhancement on ECCM capabilities.
In contrast, random PRI is an effective technique to raise the capabilities of ECCM as well as range and Doppler ambiguities suppression \cite{Filippo2001}.

In the case of the random PRI pulse train, the ambiguity function (AF) becomes a random variable and its
statistical characteristics are thus usually concerned.
Effects of PRI randomization on range and Doppler ambiguities suppression are in general studied through the AF or its expectation and variance \cite{Kaveh1976,Rasool2010,Liu2012}.
However, the conclusions therein are merely illustrated through numerical experiments and rigorous theoretical analyzes are deficient.
In this paper, we investigate the AF in an analytical manner and present some important properties on the expectation and standard deviation of AF.
The analytical results reveal the intrinsic quality of AF in the random PRI case more precisely, which contribute to designing the waveform parameters of a random PRI radar.

Researches concerning the random PRI are still ongoing \cite{Vergara-Dominguez1993,Liu2013,Chen2014}.
However, it seems that random PRI is not commonly used to control ambiguities \cite{Rasool2010}.
One major reason might be the impact of the sidelobe pedestal caused by random PRI pulses.
In situations where there exists multiple targets or heavy clutter background,
weak targets may be concealed by the aliased sidelobe floor of the strong targets or clutter.
%In order to solve this problem, we propose a signal process method to eliminate the sidelobe pedestal.
In order to solve this problem, a signal processing method based on orthogonal matching pursuit (OMP) \cite{Tropp2007} is applied to eliminate the sidelobe pedestal.
Another major drawback might be the large computation burden for processing the random PRI pulses \cite{Rasool2010}.
This problem will become a minor issue with the development of computational capability.

The remainder of the paper is organized as follows.
Section II provides the signal model.
In Section III, the statistical characteristics of the AFs are analyzed.
Section IV constructs the processing sketch for random PRI pulse train, where OMP is employed to eliminate the sidelobe pedestal.
%In Section IV, the signal processing procedure for a random PRI pulse train are proposed.
The simulation results are shown in Section V to verify the superiority and feasibility of random PRI and the OMP method.
Section VI briefly concludes this paper.

\section{Signal Model}
\label{sec:signal}
The transmitted signal with $M$ coherent pulses can be expressed as
\begin{equation}\label{equ:transmission}
	y(t) =\sum_{m=0}^{M-1} s\left(t- \sum_{k=0}^{m}T_k\right),
\end{equation}
where $T_0 = 0$ and $T_k, k = 1,2,...,M-1$, denote the intervals between the $k$th and the $(k-1)$th pulse, and the baseband envelope $s(t)$ is a rectangular pulse with pulse width $T_p$, i.e.,
\begin{equation}
   s(t) = \left\{ \begin{array}{ll}
		  1, & 0 < t \leq T_p,  \\
	      0, &  \text{else}.  \\
			\end{array} \right.
\end{equation}
For a random PRI radar, the pulse intervals $T_k,k = 1,2,...,M-1,$ vary randomly around a given value $T_r$.
We define the jitter (or bias) between the starting time of the $n$th pulse and the referred time $nT_r$ as
\begin{equation}
	\varepsilon_n =\sum_{k=0}^{n}T_k- nT_r,\  n=1,2,...M-1.
\end{equation}

Assuming that the jitters $\varepsilon_n,n=1,2,...M-1,$ are independent and identically distributed (i.i.d.) random variables which are subject to an uniform distribution $\text{U}(-\rho/2,\rho/2)$.
In avoidance of the interlacing over the successive pulses,
 the range of jitters $\rho$ is limited to $\rho \le T_r -2T_p$.
In the specific case that $\rho \rightarrow 0$, $T_k$ equals a constant $T_r$.
The transmitted waveform then reduces to a stable PRI pulse train, i.e.,
\begin{equation}\label{equ:transmission_stable}
	x(t) =\sum_{m=0}^{M-1} s\left(t- mT_r\right).
\end{equation}

\begin{figure}[!t]
\centering
\subfloat[$|\Lambda_{xx}(\tau,0)|$ and $|\Lambda_{yy}(\tau,0)|$]{\includegraphics[width=1.64in]{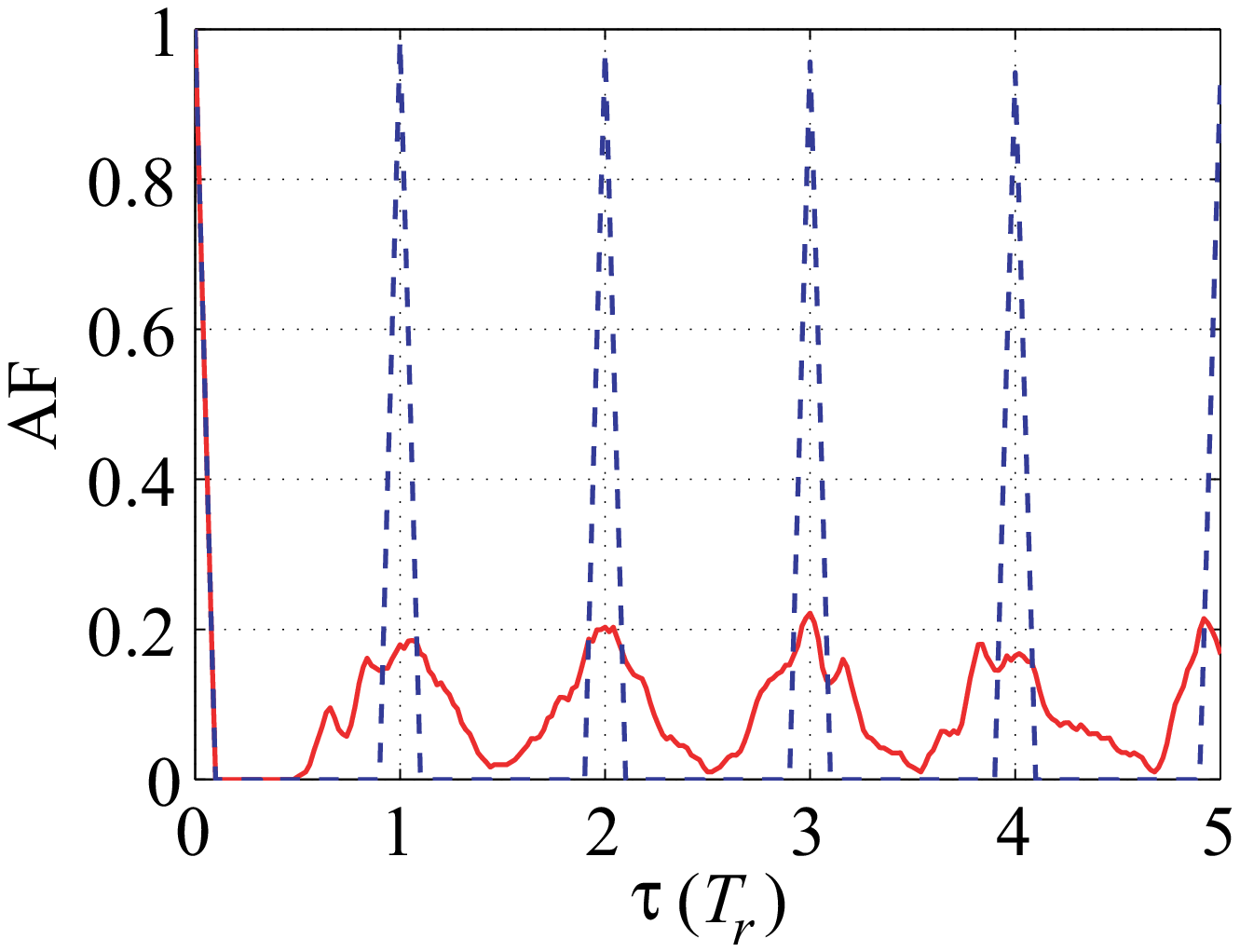}}
\hfil
\subfloat[$|\Lambda_{xx}(0,f)|$ and $|\Lambda_{yy}(0,f)|$]{\includegraphics[width=1.64in]{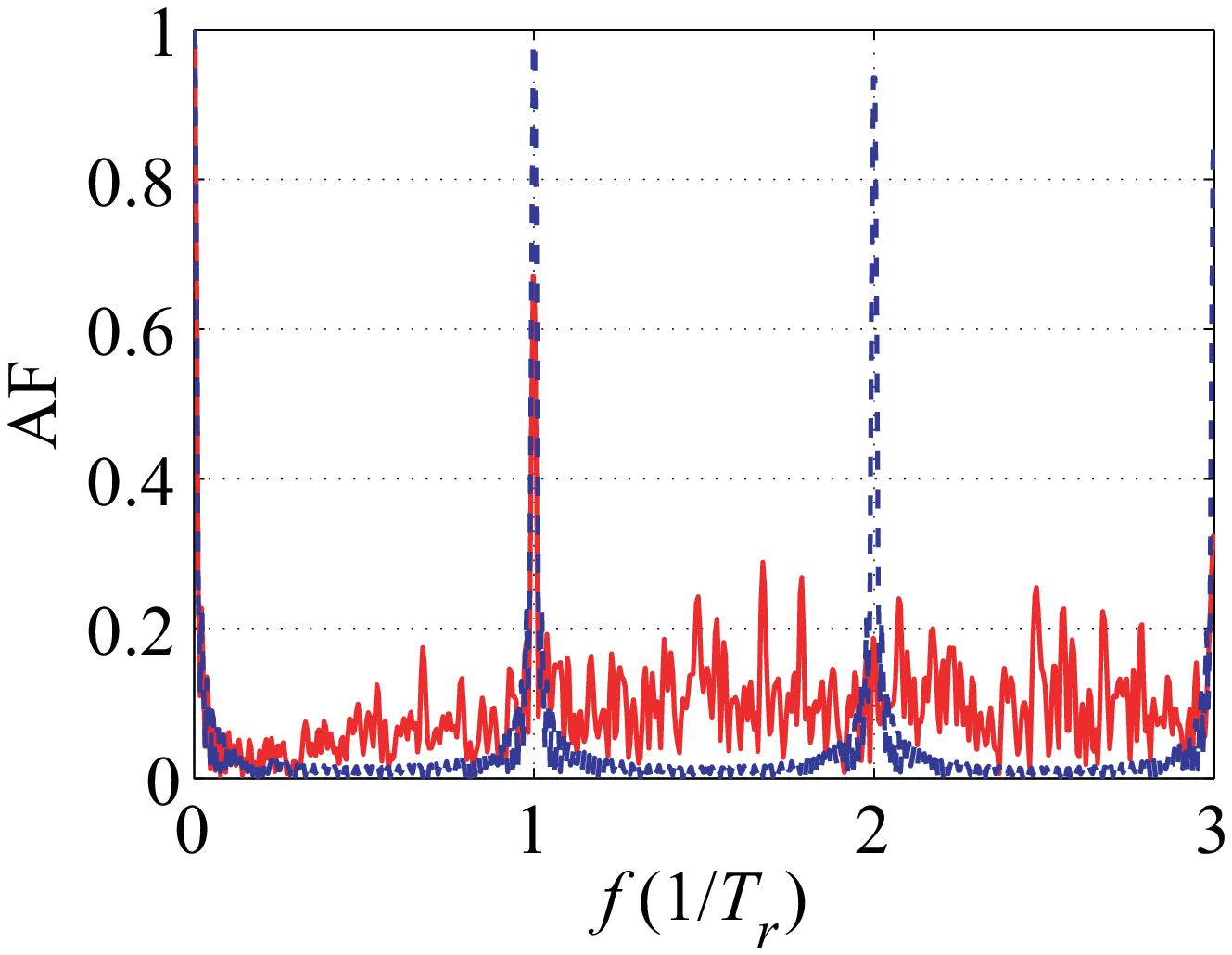}}
\caption{AFs of a stable PRI waveform (blue dashed) and a random PRI waveform (red solid).}
\label{fig:random_example}
\end{figure}
AF delineates the responses of matched filters and is usually adopted to evaluate the range-Doppler estimation performance of a waveform.
A general form of AF \cite{Levanon2004} is defined as
\begin{equation} {\label{equ:definition_AF}}
\Lambda_{uu}(\tau,f) = \int_{-\infty}^{+\infty}u(t)u^{\ast}(t-\tau)e^{-j2\pi ft}dt.
\end{equation}
Fig. \ref{fig:random_example} compares the AFs of a stable PRI waveform $x(t)$ with a random PRI waveform $y(t)$.
As shown in Fig. \ref{fig:random_example}, grating lobes appear \cite{Mahafza2004} at $\tau = pT_r$ and $f=q/T_r$, $p,q =\pm 1,\pm 2,...,\pm (M-1)$, in the AF of $x(t)$, which are range and Doppler ambiguities.
For a random PRI pulse train, owing to the randomness of the pulse intervals, the grating lobes are lowered. However, the volume squeezed out of the grating lobes is spreaded over the two-dimensional delay-Doppler domain, which turns into the sidelobe pedestal.

\section{Statistical Characteristics of AF}
\label{sec:statistical}

The present section concerns the statistical characteristics of the AF corresponding to the random PRI waveform.
It is a tedious work to directly manipulate the two-dimension AF $|\Lambda_{yy}(\tau,f)|$,
and notice that $|\Lambda_{yy}(\tau,f)| \le |\Lambda_{yy}(\tau,0)|$ always holds, hence the two main cross-sections are evaluated instead.
To be specific, we focus on the depiction of the expectation and standard deviation of $|\Lambda_{yy}(\tau,0)|$ and $|\Lambda_{yy}(0,f)|$, which reflect the average level and fluctuation intensity of the sidelobes in two directions, respectively.
Due to limited space, all the proofs are left to the subsequent journal paper.

\subsection{AF versus Reference Delay}
\label{ssec:ambfundelay}

\begin{lem}
$E \left[\left | \Lambda_{yy}(\tau,0) \right| \right]$ is an even function with respect to $\tau$, i.e.,
$E \left[\left | \Lambda_{yy}(\tau,0) \right| \right] = E \left[\left | \Lambda_{yy}(-\tau,0) \right| \right]$.
\end{lem}

\begin{lem}[Local monotonicity] \label{lem:monotonicity}
$E \left[\left | \Lambda_{yy}(\tau,0) \right| \right](\tau>0)$ is monotonically increasing on interval $(c_m,mT_r)$, and is monotonically decreasing on interval $(mT_r,c_{m+1})$,
where $c_m$ is a constant within $((m-1)T_r,mT_r),m=1,2,...,M-1$.
\end{lem}

\begin{pro}[Non-randomness of main lobe]
The main lobe of $|\Lambda_{yy}(\tau,0)|$ does not vary randomly, i.e.,
\begin{align}
|\Lambda_{yy}(\tau,0)| = |\Lambda_{xx}(\tau,0)|, 0\le \tau \le T_p.
\end{align}
\end{pro}

\begin{thm}[Peaks of sidelobes] \label{thm:range_peak}
The values of normalized AF $G(\tau):= E \left[\left | \Lambda_{yy}(\tau,0) \right| /\left| \Lambda_{yy}(0,0) \right|\right](\tau \ne 0)$ at local maxima $\tau = pT_r,p = \pm 1,\pm2,...,\pm(M-1)$, are given by
\begin{align} {\label{equ:range_sidelobe_peak}}
G(pT_r) =%E \left[| \Lambda_{yy}(pT_r,0)| \right] =
 \left\{ \begin{array}{ll}
		  \displaystyle\frac {M-|p|}{M}  (1-\frac{\rho}{3T_p}), & \rho < T_p, \\
	      \displaystyle\frac {M-|p|}{M}  (\frac{T_p}{\rho}-\frac{T_p^2}{3\rho^2}),&  \rho > T_p. \\
			\end{array} \right.
\end{align}
\end{thm}

\begin{thm}[Standard deviation of peaks] \label{thm:range_peak_std}
The standard deviation of $\left| \Lambda_{yy}(pT_r,0) \right|/\left| \Lambda_{yy}(0,0) \right|$, $p=\pm 1,\pm 2,...,\pm (M-1)$ satisfies
\begin{align} {\label{equ:standard_deviation_for_range}}
{\rm{std}} \left[\frac{\left|\Lambda_{yy}(pT_r,0)\right|}{\left|\Lambda_{yy}(0,0)\right|}\right]  \le
\displaystyle\frac{\sqrt{M-|p|}}{M} \sqrt{\frac{2T_p}{3\rho}}, \rho > T_p.
\end{align}
\end{thm}

{\bf{Remarks}}.
For a stable PRI waveform, the grating lobes are located at $\tau = pT_r, p=\pm1,\pm2,...,\pm(M-1),$ with relative peak values $(M-|p|)/M$.
Compared with the stable PRI case, the range ambiguities of random PRI are suppressed and the sidelobe pedestal is formed due to the dispersed volume of grating lobes based on Lemma \ref{lem:monotonicity} and Theorem \ref{thm:range_peak}.
It is also known from Theorem \ref{thm:range_peak} that the average range sidelobe level depends on the range of jitters $\rho$,
and becomes lower with the increase of $\rho$.
The variation of $\left | \Lambda_{yy}(pT_r,0) \right|$ is subject to the number of the pulses $M$ as revealed by Theorem \ref{thm:range_peak_std},
and ${\rm{std}} \left[\left | \Lambda_{yy}(pT_r,0) \right| /\left | \Lambda_{yy}(0,0) \right| \right] < 1/\sqrt{M}$ holds.

Theorems \ref{thm:range_peak} and \ref{thm:range_peak_std} are useful to guide the design of waveform parameters for random PRI to achieve the desired range sidelobe level.
For instance, in order to achieve a normalized range sidelobe level lower than $0.25$, $\rho=5T_p$ and $M\ge54$ can be chosen to ensure $G(pT_r)<0.2$ and ${\rm{std}} \left[\left | \Lambda_{yy}(pT_r,0) \right|/\left | \Lambda_{yy}(0,0) \right| \right] < 0.05$, which implies $G(pT_r) + {\rm{std}}\left[\left | \Lambda_{yy}(pT_r,0) \right|/\left | \Lambda_{yy}(0,0) \right| \right]<0.25$. Then
the normalized range sidelobe level is less than $0.25$ with a large probability.

\subsection{AF versus Reference {D}oppler}
\label{ssec:ambfundoppler}

The expectation of the normalized AF versus reference Doppler $B(f):= E \left[\left | \Lambda_{yy}(0,f) \right| /\left| \Lambda_{yy}(0,0) \right|\right]$ can be expressed as
\begin{align} \label{equ:doppler_ambiguity}
& \ \ \ \ \ \ \ \ \ \ \ \ \ \ \ \ \ \ \ \ B(f) = \frac{\left|{\rm{sinc}} (\pi f T_p)\right|}{M} \times \\
& E \left[\Big( M +  2 \sum_{p=1}^{M-1} \sum_{n=0}^{M-p-1} \cos (2\pi f(pT_r -\varepsilon_{n+p}+\varepsilon_{n}) )\Big)^{\frac{1}{2}} \right]. \nonumber
\end{align}
The closed-form of (\ref{equ:doppler_ambiguity}) is hard to obtain;
therefore, an upper bound $B_u(f)=E\left[|\Lambda_{yy}(0,f)|^2\right]^{1/2}/|\Lambda_{yy}(0,0)|$ and a lower bound $B_l(f)=\left|E\left[\Lambda_{yy}(0,f)\right] \right|/|\Lambda_{yy}(0,0)|$ are derived to study the statistical characteristics of $|\Lambda_{yy}(0,f)|$.

%To facilitate the analyses, a pair of upper and lower bounds of $B(f)$ are proposed in the following theorem.
\begin{thm}[Upper and lower bound] \label{thm:doppler_bound}
The upper bound and lower bound of $E\left[|\Lambda_{yy}(0,f)|/|\Lambda_{yy}(0,0)| \right]$ are
\begin{align}
  & \ \ \ \ \ \ \ \ \ \ \ \ \ \  \ \ \ \ \ \ \   B_u(f) = | {\rm{sinc}}(\pi f T_p)|  \times  \\
 &   \ \ \ \ \ \    \left(\frac{1}{M} + \left(\left|\frac{\sin( \pi M f T_r)}{M\sin (\pi f T_r)}\right|^2 - \frac{1}{M}\right)  {\rm{sinc}}^2(\pi  f\rho)\right)^{\frac{1}{2}}, \nonumber \\
 & \ \ \  B_l(f)= \frac{\left|{\rm{sinc}} (\pi f T_p) {\rm{sinc}}(\pi f\rho)\right|}{M} \times \left|\frac{\sin( \pi MfT_r)}{\sin( \pi fT_r)}\right|.
\end{align}
\end{thm}

\begin{thm}[Peak of sidelobes] \label{thm:doppler_peak}
Denote $\mathscr{E}$ as the set of all the nonzero local maxima for $B_u(f)$.
%Denote $\mathscr{E}$ as the set of the local maximizers of $B_u(f)$ for $f \ne 0$.
%Let $\mathscr{E}$ collect the local maximizers of $B_u(f)$, $f \neq 0$.
Then, the maximum of $B_u(f)$ on $\mathscr{E}$ is constrained as
\begin{align} {\label{equ:doppler_sidelobe_peak}}
\max\limits_{f\in\mathscr{E}} B_u(f) < \sqrt{\frac{1}{M} + (1 - \frac{1}{M}){\rm{sinc}}^2(\pi w)},
\end{align}
where
\begin{align}
w = \min\left(w^*,\frac{(M-1)\rho}{MT_r}\right),
\end{align}
\begin{align}
w^*=\inf\left\{w>0\left|\text{sinc}^2(\pi w)=\text{sinc}^2(\pi w_0)\right.\right\},
\end{align}
and $w_0$ is the maximum for $\text{sinc}^2(\pi w)$ on interval $w\in(1,2)$.
\end{thm}

\begin{thm}[Standard deviation] \label{thm:doppler_standard deviation}
The standard deviation of $\left|\Lambda_{yy}(0,f)\right|/\left|\Lambda_{yy}(0,0)\right|$ satisfies
\begin{align}
    {\rm{std}} \left[\frac{\left|\Lambda_{yy}(0,f)\right|}{\left|\Lambda_{yy}(0,0)\right|}\right] \le
    \sqrt{\frac{1-{\rm{sinc}}^2(\pi f \rho)}{M}}| {\rm{sinc}}(\pi f T_p)|.
\end{align}
\end{thm}

\begin{cor}[Non-randomness of main lobe]
If $M$ is sufficiently large, then the main lobe of $\left|\Lambda_{yy}(0,f)\right|$ approximates to the counterpart of $\left|\Lambda_{xx}(0,f)\right|$
\begin{align}
E\left[\frac{|\Lambda_{yy}(0,f)| }{|\Lambda_{yy}(0,0)|}\right]\approx  \frac{|\Lambda_{xx}(0,f)| }{|\Lambda_{xx}(0,0)|}, |f|<\frac{1}{MT_r},
\end{align}
\begin{align}
    {\rm{std}} \left[\frac{\left|\Lambda_{yy}(0,f)\right|}{\left|\Lambda_{yy}(0,0)\right|}\right] \approx 0, |f|<\frac{1}{MT_r}.
\end{align}
\end{cor}

{\bf{Remarks}}.
Theorem \ref{thm:doppler_peak} indicates that the amplitudes of Doppler sidelobes and grate lobes can be lowered by increasing the number of pulse $M$ and the range of jitters $\rho$.
Meanwhile, we can learn from Theorem \ref{thm:doppler_standard deviation}
%From Theorem \ref{thm:doppler_standard deviation}, we know
that the variation of $\left | \Lambda_{yy}(0,f) \right|$ decreases with increasing the number of pulses $M$ and satisfies ${\rm{std}} \left[\left | \Lambda_{yy}(0,f) \right| /\left | \Lambda_{yy}(0,0) \right| \right] < 1/\sqrt{M}$.

Theorems \ref{thm:doppler_peak} and \ref{thm:doppler_standard deviation} can be applied to design the waveform parameters of random PRI to acquire the desired Doppler sidelobe level.
For instance, if a normalized Doppler sidelobe level lower than $0.25$ is required,
then, $M=256$ and $\rho\ge 0.85T_r$ can be selected to ensure that the normalized standard deviation ${\rm{std}}\left[\left | \Lambda_{yy}(0,f) \right|/\left | \Lambda_{yy}(0,0) \right| \right]$ is less than $0.0625$ and the maximum of $B_u(f)$ on $\mathscr{E}$ is lower than $0.18$, such that it obeys $B(f)+{\rm{std}}\left[\left | \Lambda_{yy}(0,f) \right|/\left | \Lambda_{yy}(0,0) \right| \right] < 0.25$ for $|f|>1/(MT_r)$. Hence the requirement of Doppler sidelobe level can be realized with a large probability.

\section{Signal Processing}
\label{sec:sigpro}

Matched filtering is a key ingredient in conventional radar signal processing,
where the signal-to-noise ratio (SNR) is maximized.
For a random PRI waveform, the pulses should be aligned in range filtering and discrete Fourier transformation (DFT) is applied to Doppler filtering.
Based on the output of matched filtering,
%the targets' parameters can be obtained through peaks detection.
the parameters of the targets can be obtained through peaks detection.
The aforementioned DFT-based moving target detection method is noted as DFT-MTD.

Recall that the random PRI waveforms can effectively suppress the range and Doppler ambiguities and enhance ECCM capacities, which are the major predominances compared to the stable PRI waveforms.
However, the sidelobe pedestal emerges due to the removed volume of grating lobes.
The weak targets could be concealed by the aliased sidelobe floor of the strong targets or the heavy clutter
if DFT-MTD is adopted for processing random PRI waveforms.
In order to recover the echoes of weak targets, the impact of the strong targets and the heavy clutter should be removed.
OMP \cite{Tropp2007} is an efficient compressed sensing algorithm, which successively detects targets and eliminates the corresponding echoes.
A key step is to apply orthogonal projection filtering to thoroughly remove the returns from the detected target, which can also be found in \cite{Zhang2004,Liu2008}.

Assume that $\alpha _k$, $\tau _k$ and $f_k$ are the unknown amplitude, time delay, and Doppler frequency of the $k$th target, respectively.
Sampled at a rate of $1/T_s$, the contaminated echo from $K$ targets is given by
\begin{equation}
\begin{array}{c}
z(nT_s)= \displaystyle \sum_{k=1}^{K} \alpha_{k} y(nT_s-\tau_k) e^{j2\pi f_k nT_s} + n_0(nT_s),\\
\ n = 0,1,2,...,N-1,
\end{array}
%\nonumber\\\ \ \ \ \ \ \ \ \ \ \ \ \ \
\end{equation}
where $n_0$ is a complex Gaussian white noise with a variance of $\sigma^2$ and $N$ is the number of samples.
The received samples can also be written in a form of matrix
\begin{align}
\bm{z} =  \bm{S}\bm{a} + \bm{n}_0,
\end{align}
where $\bm{z}=[z(0),z(T_s),...,z((N-1)T_s)]^T$,
$\bm{a} = [\alpha_1,\alpha_2,...,\alpha_K]^T$, $\bm{n}_0=[n_0(0),n_0(T_s),...,n_0((N-1)T_s)]^T$,
$\bm{S}=[\bm{s}(\tau_1,f_1),\bm{s}(\tau_2,f_2),...,\bm{s}(\tau_K,f_K)]$,
and
$\bm{s}(\tau_k,f_k) = [y(-\tau_k),y(T_s-\tau_k) e^{j2\pi f_k T_s},...,y((N-1)T_s-\tau_k) e^{j2\pi f_k (N-1)T_s}]^T$, $k=1,2,...,K$.

\begin{table}[!t]
\renewcommand{\arraystretch}{1.3}
\caption{Signal Processing Procedures \cite{Tropp2007}}
\label{tab:OPF}
\centering
\begin{tabular}{p{8.2cm}}
\hline
1) $k=1,\bm{z}^{(0)}=\bm{z}$.\\
2) Apply matched filtering to data $\bm{z}^{(k-1)}$, then the estimated parameters of the $k$th target $\{\hat{\tau_k},\hat{f_k}\}$ are obtained through peak detection,i.e.,
\begin{equation}
 \{\hat{\tau_k},\hat{f_k}\} = {\rm{arg}} \max\limits_{\{\tau_k,f_k\}} \{ | \bm{s}^H( \tau_k,f_k) \bm{z}^{(k-1)} |\},
\end{equation}
where $(\cdot)^H$ denotes the conjugate transpose operation.\\
3) Reconstruct the steering vector of the $k$th detected target as
$\bm{s}(\hat{\tau}_k,\hat{f}_k) $.
The orthogonal projection matrix is updated
\begin{equation}
 %\bm{P}_{\bm{A}}^{\bot} = \bm{I} - \bm{A}^{(k)}(\bm{A}^{(k)h}\bm{A}^{(k)})^{-1}\bm{A}^{(k)H},
 \bm{P}_{\bm{A}}^{\bot} = \bm{I} - \bm{A}(\bm{A}^H \bm{A})^{-1}\bm{A}^{H},
\end{equation}
where $\bm{A}=[\bm{s}( \hat{\tau}_1,\hat{f}_1),\bm{s}( \hat{\tau}_2,\hat{f}_2),...,\bm{s}( \hat{\tau}_k,\hat{f}_k)]$.\\
4) Apply orthogonal projection filtering to data $\bm{z}$
\begin{equation}
 \bm{z}^{(k)}= \bm{P}_{\bm{A}}^{\bot}\bm{z}.
\end{equation}
5) $k=k+1$. If $k$ exceeds the num of the targets, the signal is terminated, otherwise return to 2).\\
\hline
\end{tabular}
\end{table}
The detailed signal processing method is shown in Table \ref{tab:OPF}, where the number of the targets can be estimated according to references \cite{Schwarz1978,Schmidt1986,Wu1995}.
Compared with the DFT-MTD method, the simultaneous detection and parameters estimation for all targets is avoided. Instead, a series of detections and parameters estimations are performed, each of which corresponds to a designated target.
Consequently, the weak target detection problem is solved.

\section{Numerical Simulations}
\label{sec:simulation}

\subsection{Statistical Characteristics of AF}
\begin{figure}[!t]
\centering
\subfloat[$\rho =0$]{\includegraphics[width=1.64in]{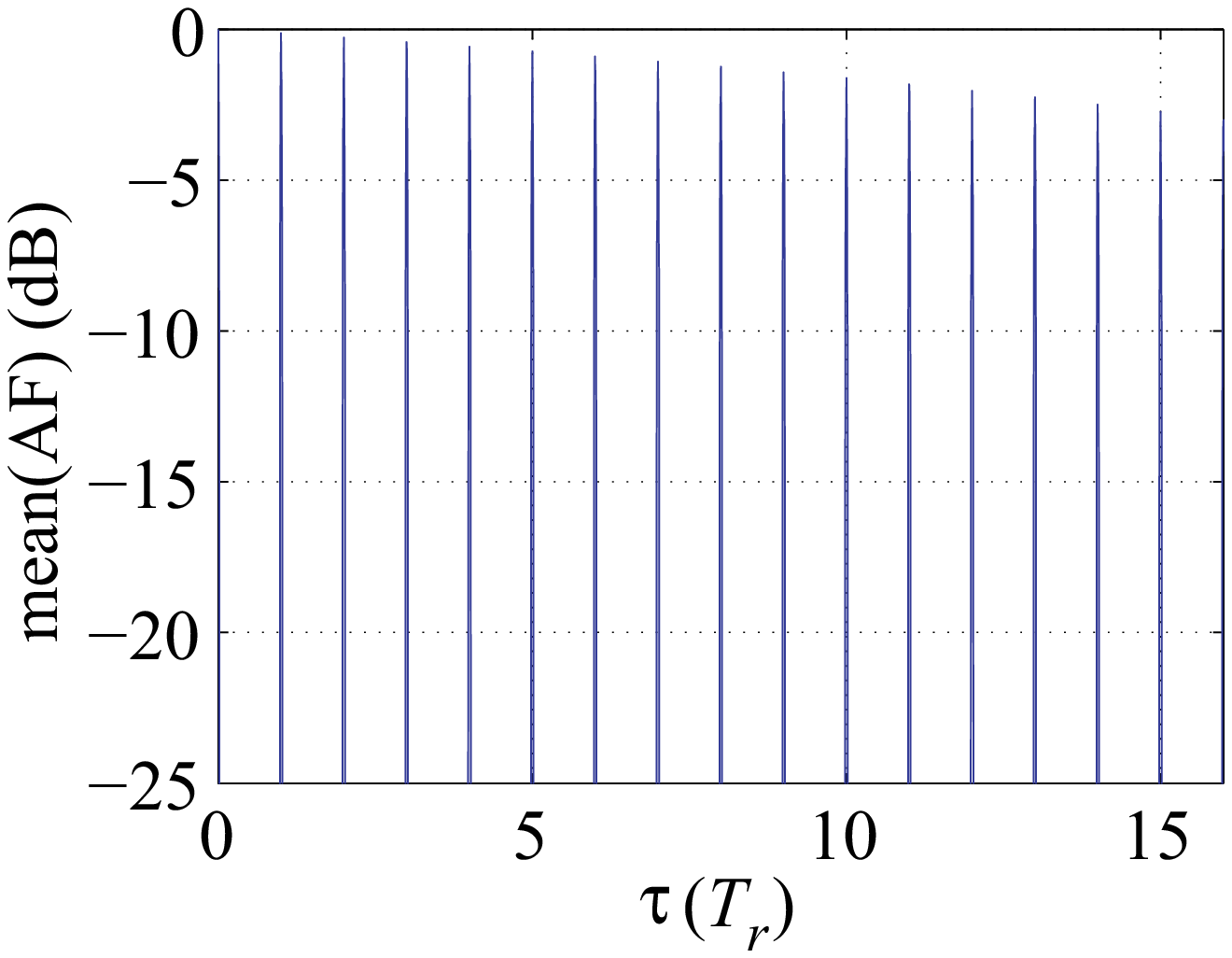}}
\hfil
\subfloat[$\rho =0.1T_r$]{\includegraphics[width=1.64in]{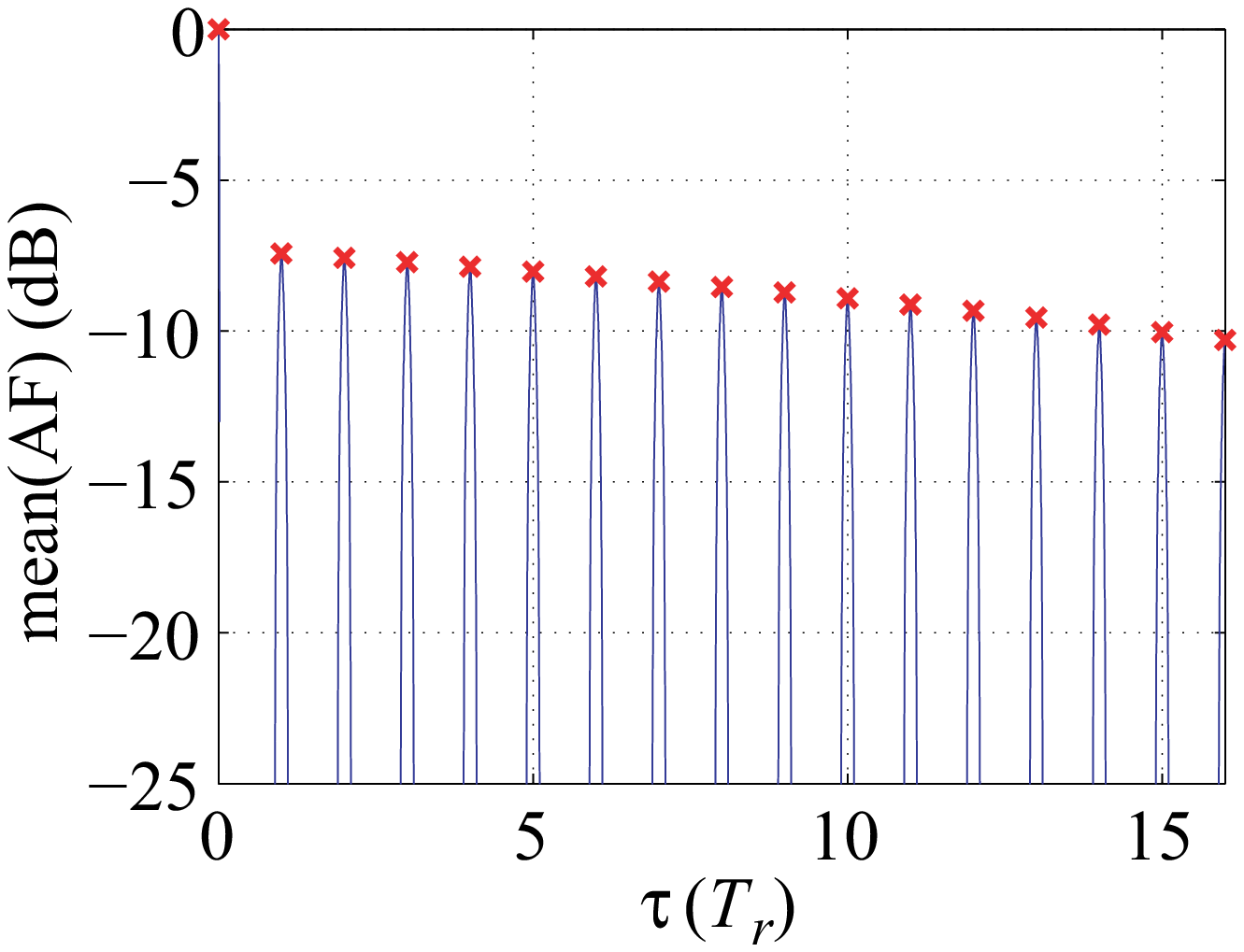}}
\vfil
\subfloat[$\rho =0.5T_r$]{\includegraphics[width=1.64in]{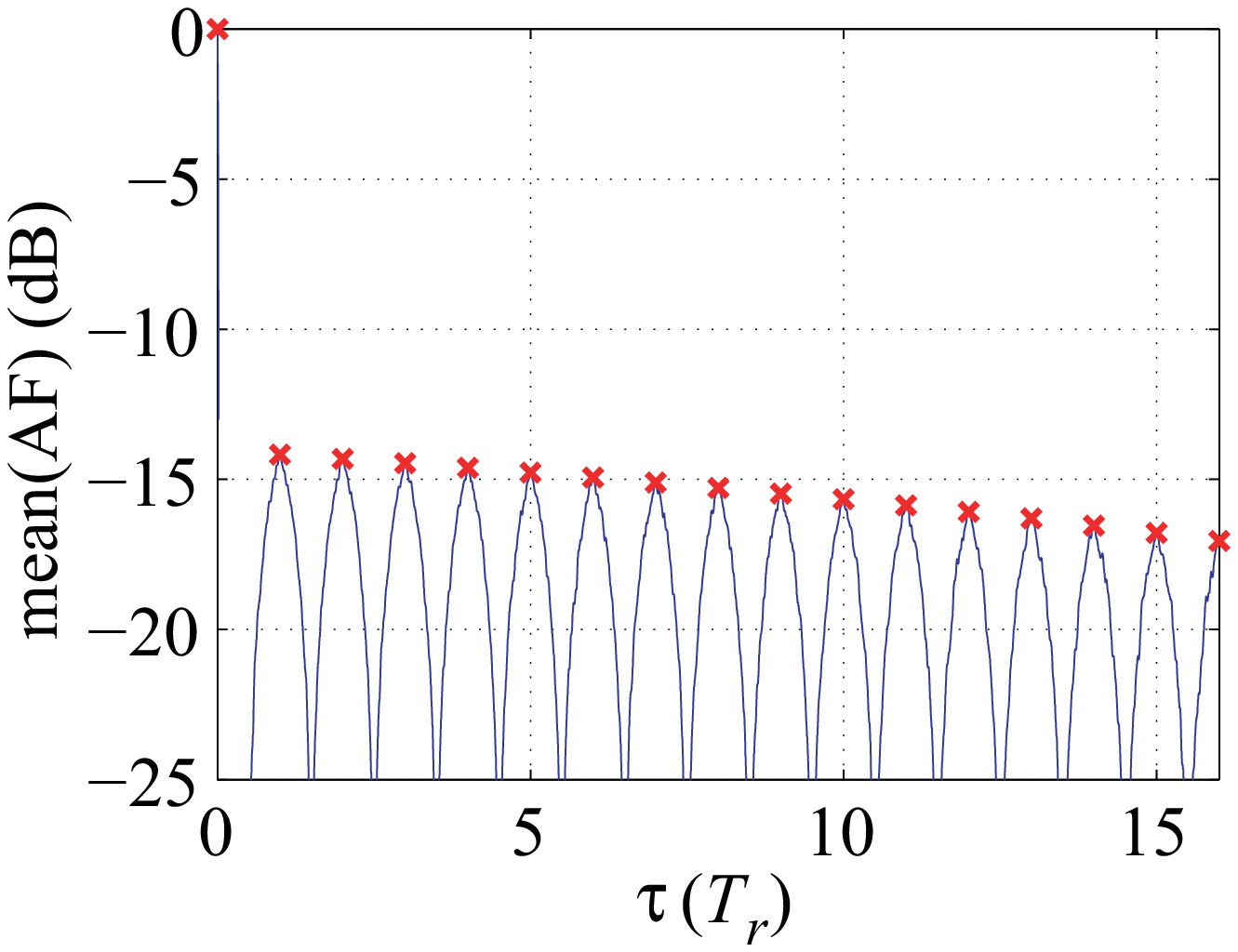}}
\hfil
\subfloat[$\rho =0.5T_r$]{\includegraphics[width=1.64in]{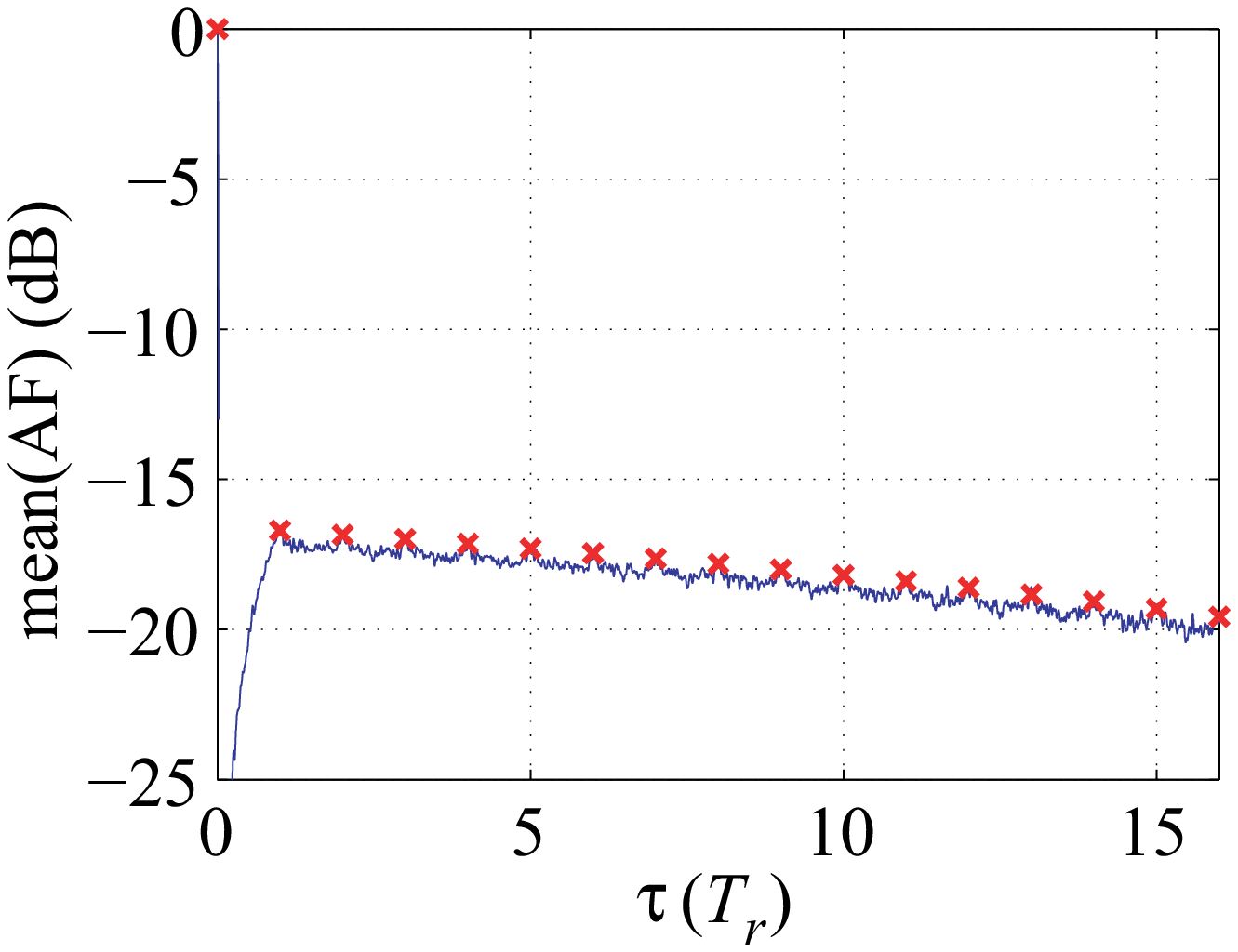}}
\caption{Simulation results of $E \left[| \Lambda_{yy}(\tau,0)|/| \Lambda_{yy}(0,0)| \right]$  (blue solid) and the theoretical values of ({\ref{equ:range_sidelobe_peak}}) (red cross).}
\label{fig:simulation_E_f_0}
\end{figure}

\begin{figure}[!t]
\centering
\subfloat[$\rho =0$]{\includegraphics[width=1.64in]{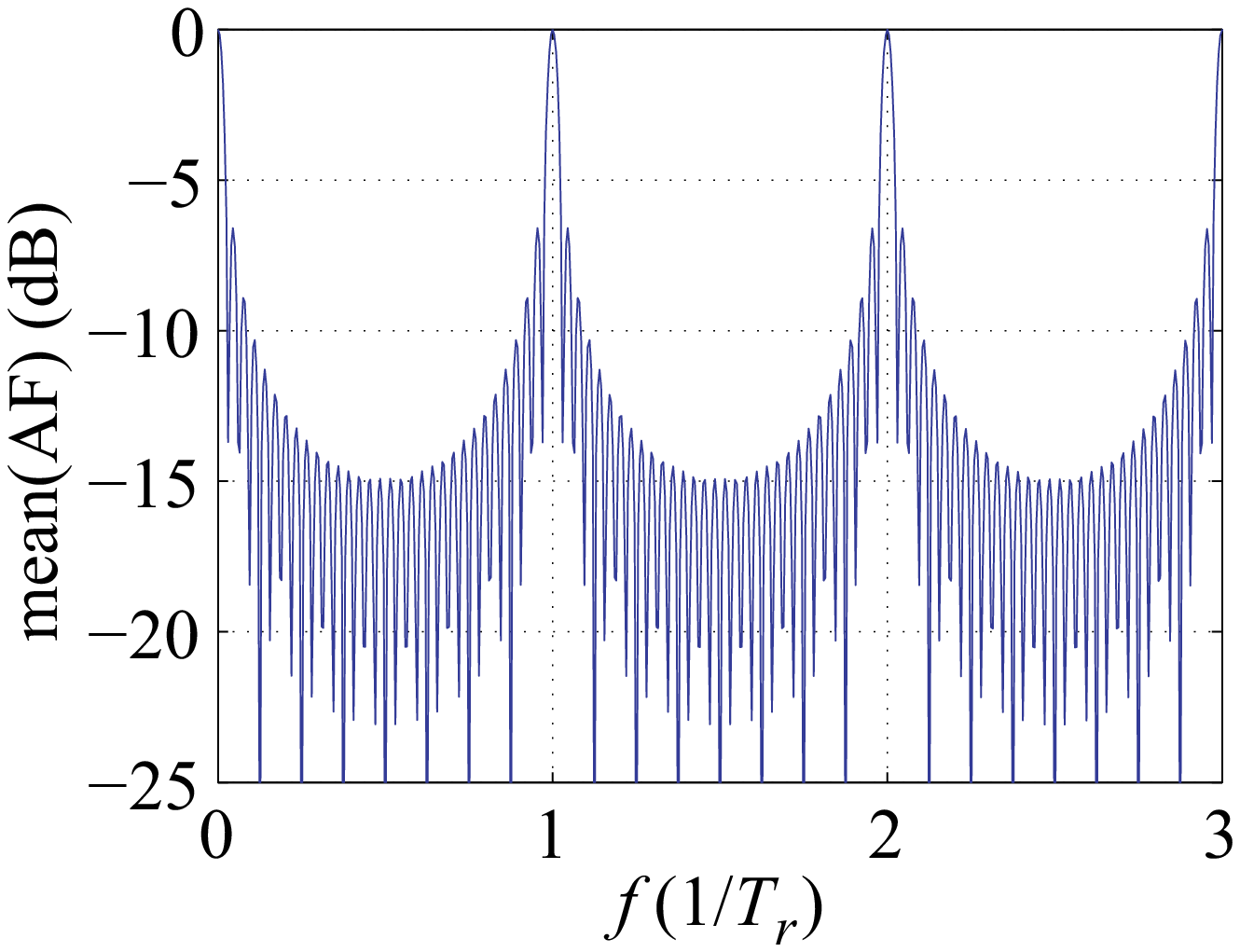}}
\hfil
\subfloat[$\rho =0.1T_r$]{\includegraphics[width=1.64in]{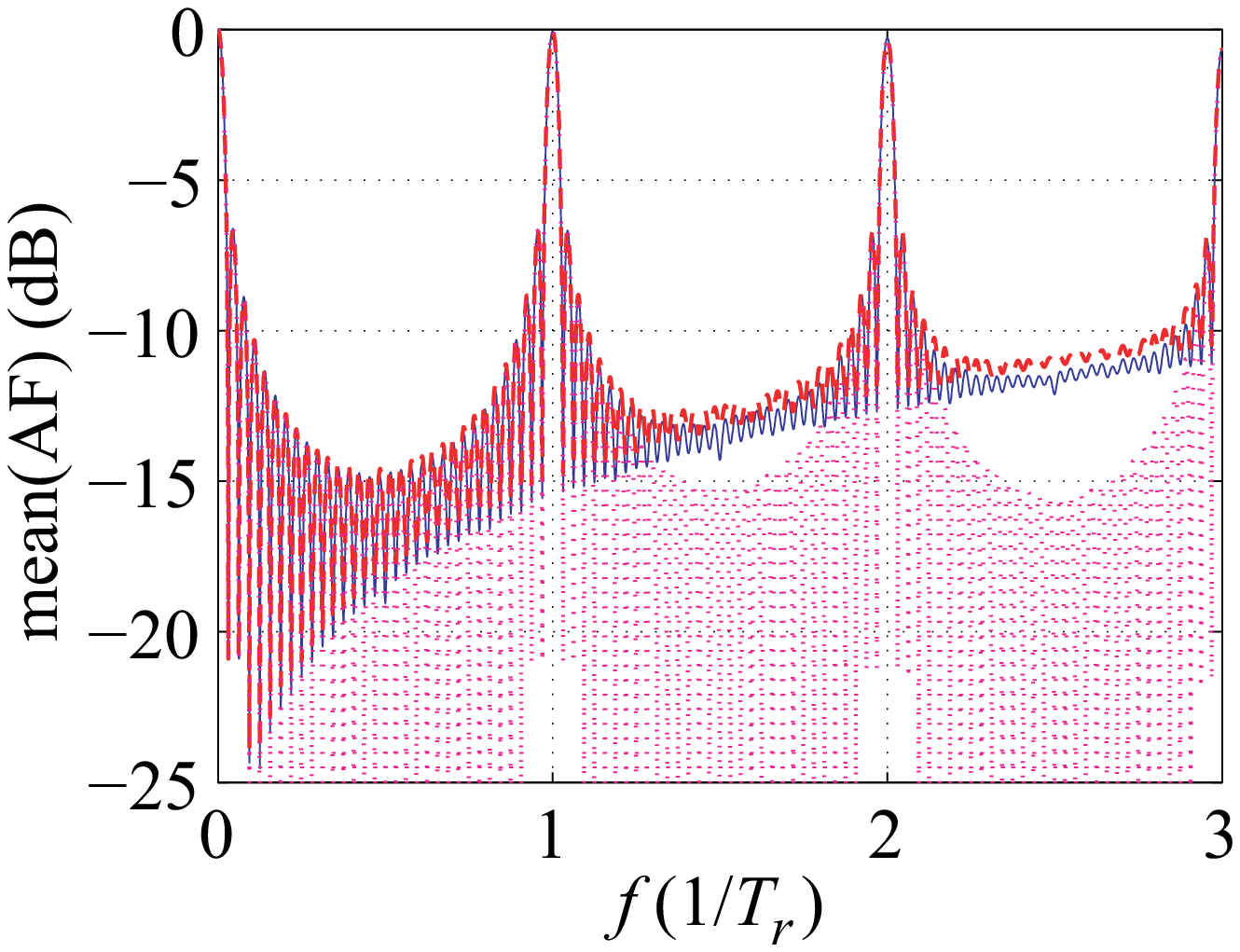}}
\vfil
\subfloat[$\rho =0.5T_r$]{\includegraphics[width=1.64in]{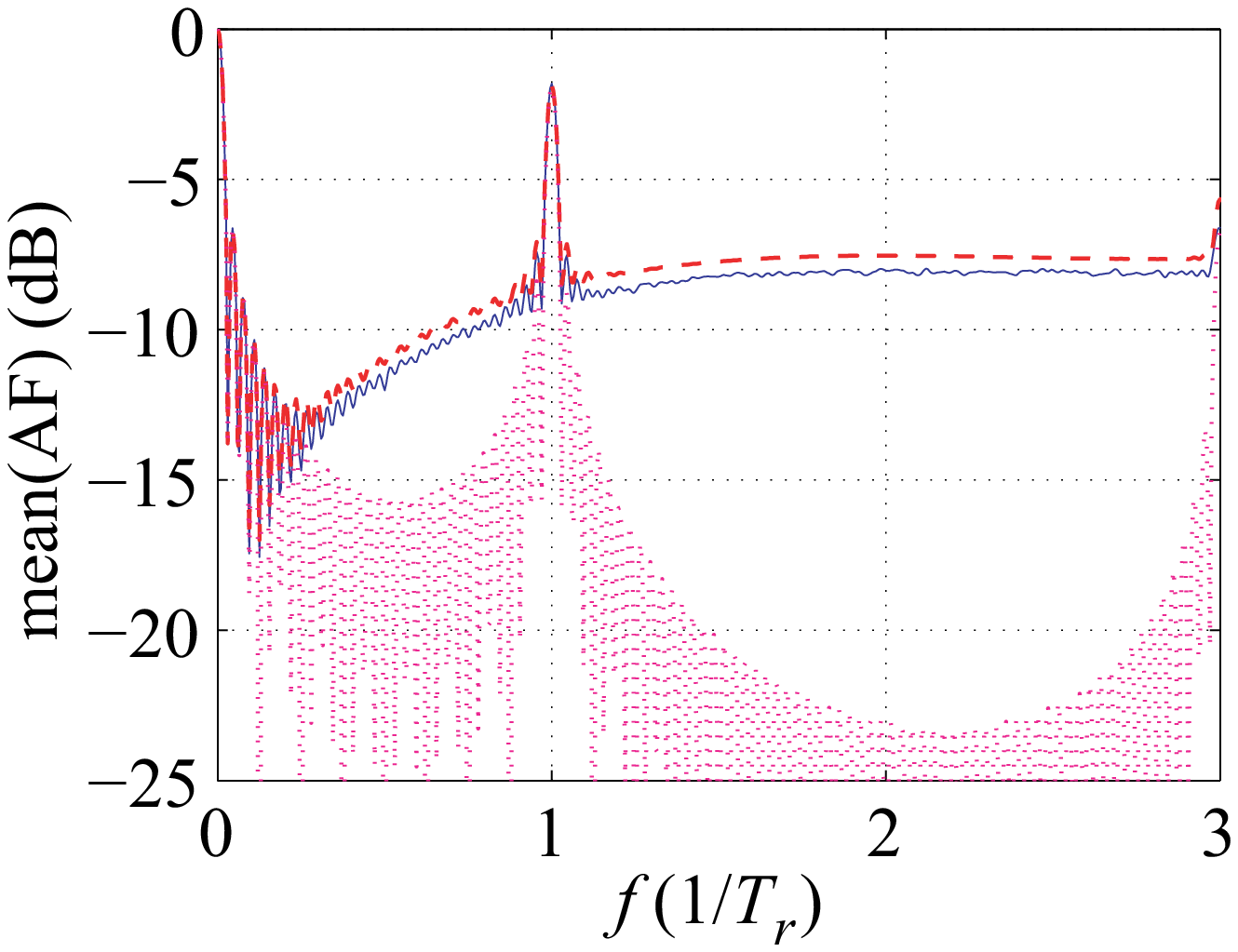}}
\hfil
\subfloat[$\rho =0.9T_r$]{\includegraphics[width=1.64in]{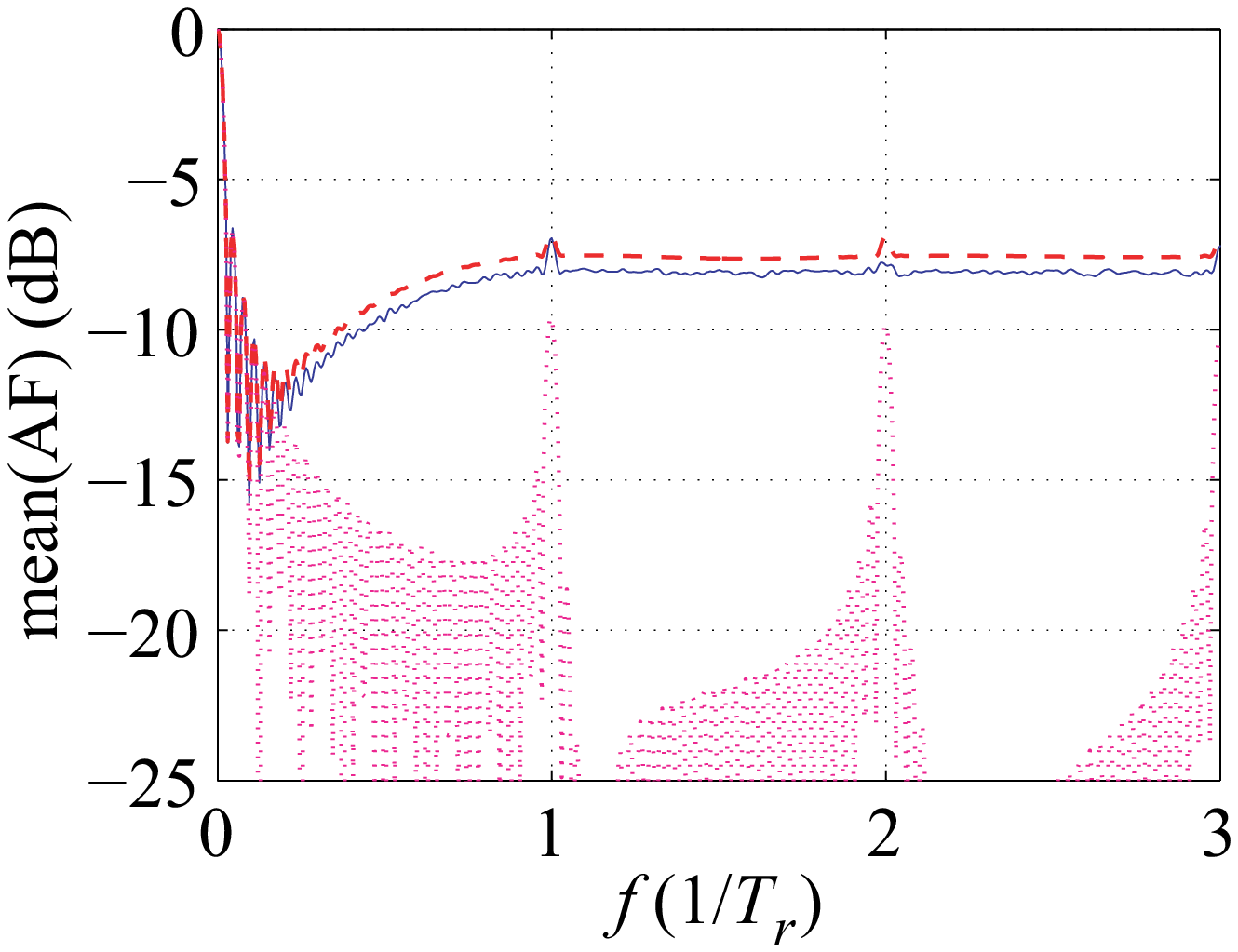}}
\caption{Simulation results of $E \left[| \Lambda_{yy}(0,f)| /| \Lambda_{yy}(0,0)\right]$ (blue solid), its upper bound (red dashed) and lower bound (magenta dotted).}
\label{fig:simulation_E_tau_0}
\end{figure}

In this subsection, we present the impact of random jitters on the statical characteristics of AFs through numerical experiments.
The parameters are displayed in Table \ref{tab:parameter_set_1}.
Fig. \ref{fig:simulation_E_f_0} and Fig. \ref{fig:simulation_E_tau_0} illustrate the expectations of $| \Lambda_{y y} (\tau, 0) |$ and $| \Lambda_{y y} (0, f) |$ for different ranges of random jitters, respectively.
The calculations of expectations are provided by computer simulation of $2000$ independent Monte Carlo (MC) trials.

\begin{table}[!t]
\renewcommand{\arraystretch}{1.3}
 \caption{Radar Waveform Parameters}
 \label{tab:parameter_set_1}
\centering
\begin{tabular}{ c  c }
\hline
Parameter & Value\\
\hline
Carrier Frequency / $f_0$ & $10$GHz\\
Pulse Number / $M$ & $32$\\
Pulse Width / $T_p$ & $1$us\\
PRI / $T_r$ & $50$us\\
Sampling Frequency / $F_s$ & $1$MHz\\
\hline
\end{tabular}
\end{table}

Fig. \ref{fig:simulation_E_f_0} shows the MC-based results of the normalized AF versus reference time delay $G(\tau)$ and the theoretical values of ({\ref{equ:range_sidelobe_peak}}). It can be observed that a small $\rho$ leads to an effective suppression of range grating lobes. Apart from that, the effect of random PRI on range ambiguities suppression is better with a larger range of jitters $\rho$.
When $\rho$ approaches $T_r$, the volume of the grating lobes is spreaded over almost everywhere in the delay domain.
All the aforementioned results are in good agreement with the conclusions in Lemma \ref{lem:monotonicity} and Theorem \ref{thm:range_peak}.

The normalized AF versus reference Doppler $B(f)$ and its upper and lower bounds are shown in Fig. \ref{fig:simulation_E_tau_0}. We notice that the upper and lower bounds proposed in Section \ref{ssec:ambfundoppler} are useful.
It can be seen from Fig. \ref{fig:simulation_E_tau_0} that the Doppler grating lobes of random PRI are lower than the counterparts of stable PRI, and they can be significantly suppressed with a large $\rho$.
However, the effect of small range of jitters $\rho$ is unapparent. We note that levels of grating lobes depend on the number of pulses $M$ as well.


As observed from Figs. \ref{fig:simulation_E_f_0} and \ref{fig:simulation_E_tau_0},
for the case that $M = 32$,
in order to obtain a satisfactory suppression (lower than $-5$dB) of range and Doppler ambiguities simultaneously, a large $\rho$ (e.g., $\rho=0.9T_r$) can be adopted in random PRI waveform designing.

\subsection{Weak Target Recovery}

In this subsection, MC simulations are employed to evaluate the performance of the OMP method on weak target signal recovery, and we also involve DFT-MTD for comparison.
Consider a random PRI pulse train consisting of $M=256$ pulses. The range of random jitters $\rho = 0.8T_r$ and the rest parameters are identical to Table \ref{tab:parameter_set_1}.

Suppose that the received signal is composed of a strong target (T$_1$) and a weak target (T$_2$).
The ranges and Doppler frequencies of T$_1$ and T$_2$ are $\{R_1,f_1\}=\{11.98\rm{km},40\rm{kHz}\}$ and $\{R_2,f_2\}=\{15.05\rm{km},30\rm{kHz}\}$, respectively.
The SNR of T$_k$ is defined as SNR$_k = |\alpha_k|^2/\sigma^2,k=1,2$.
%The SNR of T$_k$ is defined as SNR$_k = |\alpha_k|^2/N_0,k=1,2$.
The range resolution and Doppler frequency resolution are defined as $\Delta R=c/(2F_s)$ and $\Delta f=1/\sum_0^{M-1}T_k$, respectively.
For a designated target, signal recovery is considered a success if the estimation errors of its range and Doppler frequency are less than $\Delta R/2$ and $\Delta f/2$, respectively.
The probability of successful recovery $P_r$ is utilized as a metric of algorithm performance.

\begin{figure}[!t]
\centering
\subfloat[$\sigma^2$ varies]{\includegraphics[width=1.64in]{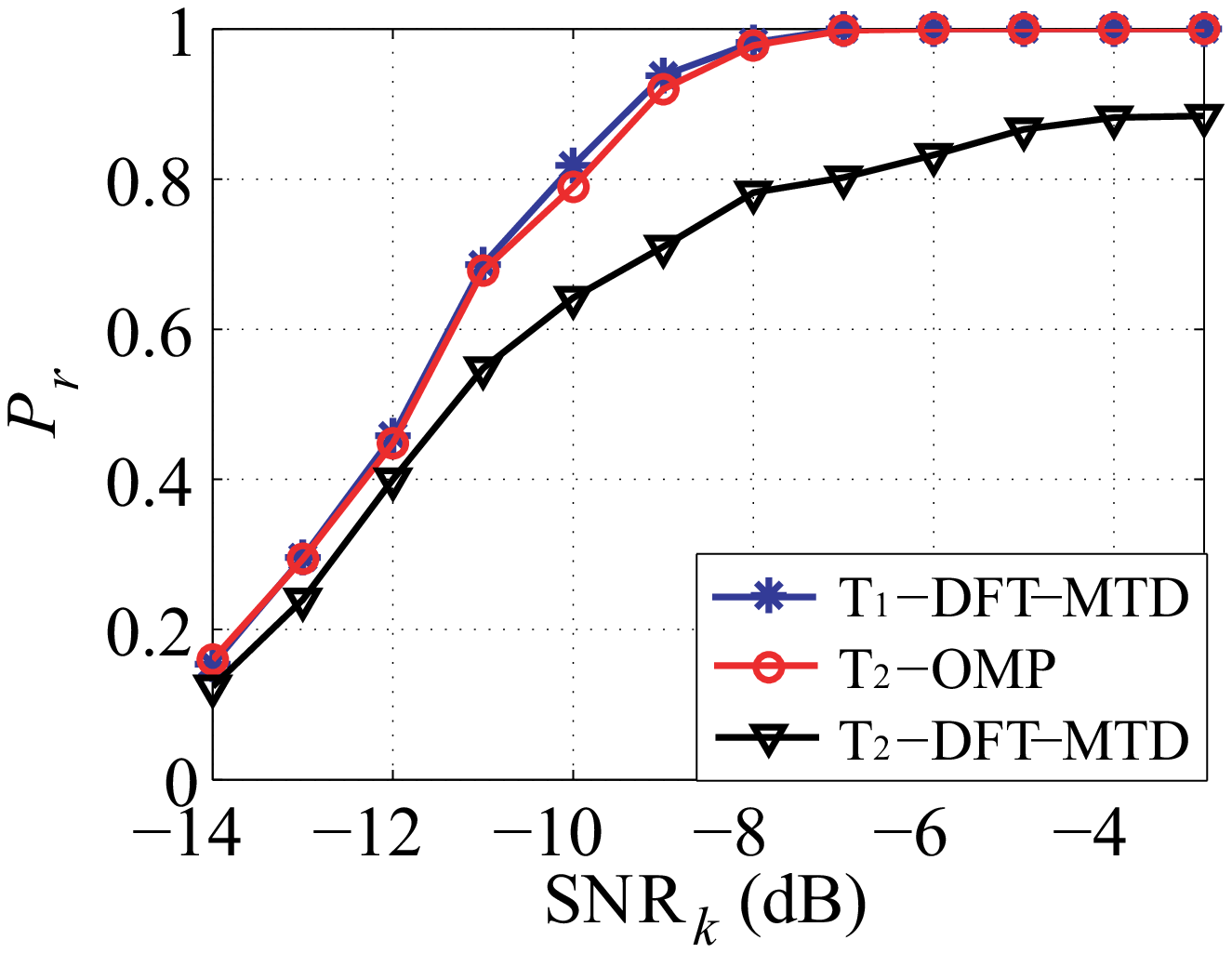}
\label{fig:rate of successful recovery_a}}
\hfil
\subfloat[$|\alpha_1|^2$ varies]{\includegraphics[width=1.64in]{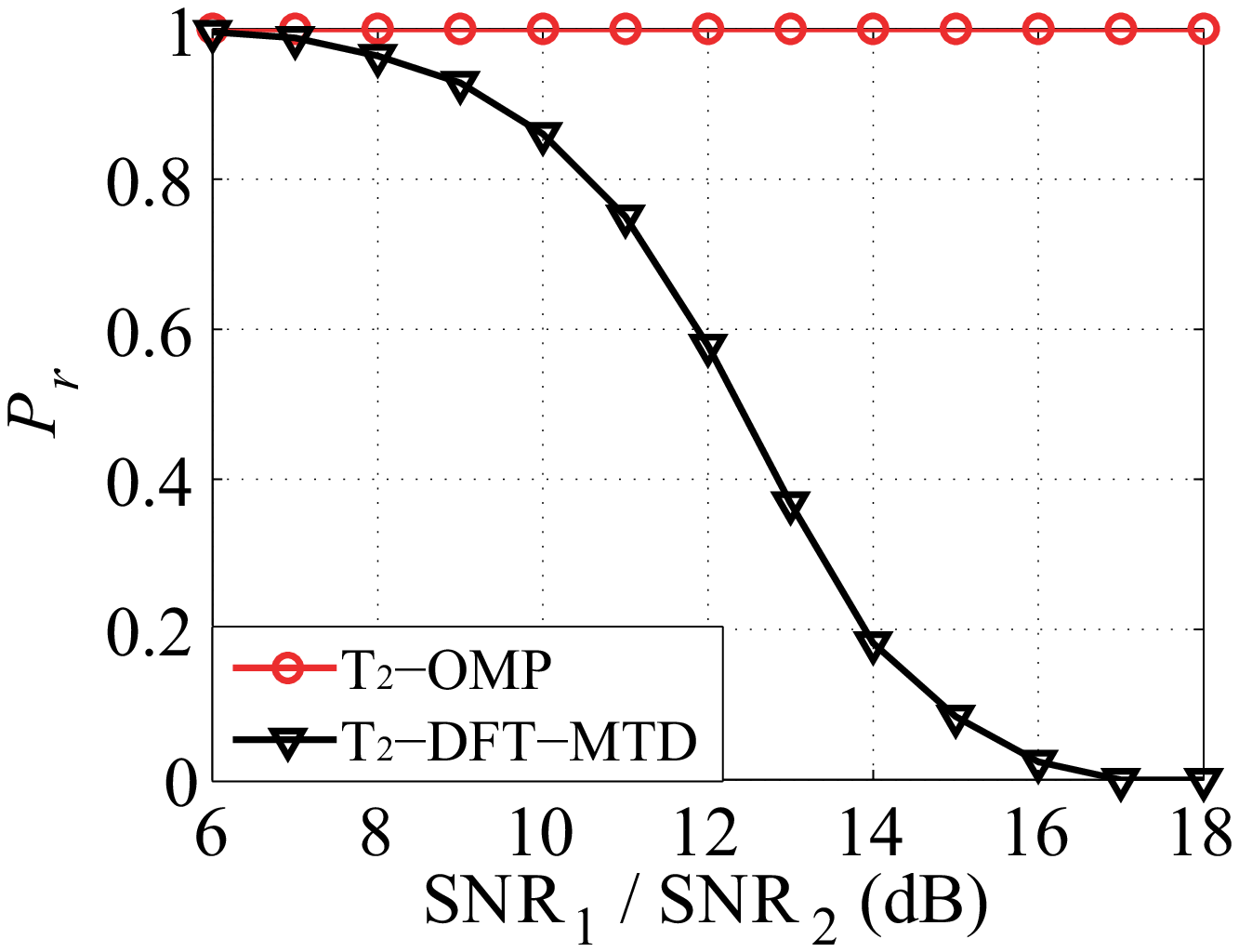}
\label{fig:rate of successful recovery_b}}
\caption{(a) $P_r$ versus SNR$_k,k=1,2$. For the sake of comparison between T$_1$-DFT-MTD and T$_2$-OMP, the abscissa are SNR$_1$ for the curve T$_1$-DFT-MTD and SNR$_2$ for the curves T$_2$-OMP and T$_2$-DFT-MTD, respectively. (b) $P_r$ versus SNR$_1$/SNR$_2$.}
\label{fig:rate of successful recovery}
\end{figure}

So as to illustrate the performances of the aforementioned two methods on weak target recovery, the following two different scenarios are employed for simulations. Within each scenario, $500$ independent MC trials are conducted. 

{\bf Scenario (a)}: both $|\alpha_1|$ and $|\alpha_2|$ are fixed and $|\alpha_1|^2/|\alpha_2|^2=10$dB, while the power of the noise $\sigma^2$ varies.
Under such a scenario, which is designed for demonstrating the impact of noise, the average performances of the two methods are shown in Fig. \ref{fig:rate of successful recovery_a}.
%The average performances of the two methods are shown in Fig. \ref{fig:rate of successful recovery_a} to demonstrate the impact of noise.
In order to compare the performances on the strong and the weak target under the same SNR rather than the same $\sigma^2$, the results of T$_1$ and T$_2$ are depicted with respect to SNR$_1$ and SNR$_2$, respectively.
Since the performance of OMP is consistent with that of DFT-MTD on T$_1$, the curve T$_1$-OMP is omitted from the figure.
Benefited from the orthogonal projecting procedure, which wipes out the impact of the strong target,
the OMP method outperforms DFT-MTD on the the weak target (i.e. T$_2$) recovery.
In addition, it is satisfying that the performance of OMP for weak target recovery is almost the same as that of DFT-MTD for strong target recovery, indicating that in terms of weak target recovery, the strong target produces trivial impact on the performance of OMP under the present scenario.

{\bf Scenario (b)}: only $|\alpha_1|$ varies, whereas $|\alpha_2|$ and $\sigma^2$ remain constant and $\text{SNR}_2=-6$dB holds. Since $\text{SNR}_2$ is fixed, the impact of the strong target T$_1$ on the performances of the two methods is explicitly revealed; see Fig. \ref{fig:rate of successful recovery_b}.
The OMP method could successfully recover the weak target, even when the power of T$_1$ is $18$dB higher than that of T$_2$, in respect that the echo from T$_1$ is removed from the received samples
and thus T$_2$ is not masked in the aliased sidelobe of T$_1$.
On the contrary, however, the performance of DFT-MTD gradually fades as the power of T$_1$ gets larger, and it almost fails to recover the weak target T$_2$ when $\text{SNR}_1$ is $16$ dB higher than $\text{SNR}_2$, %since T$_2$ is absolutely buried in the sidelobe pedestals of T$_1$.
since the sidelobe pedestal of T$_1$ is not eliminated by DFT-MTD when $T_2$ is concealed in it.

\section{CONCLUSION}
\label{sec:conclude}
In this paper, we analyzed the statistical characteristics of the AF of random PRI waveforms.
Effects of ambiguity suppression were analyzed through theoretical derivation and illustrated by simulations.
The theoretical results of the expectation and standard deviation of the AFs were crucial for the waveform design.
OMP method was applied to deal with the problem that the weak targets might be submerged in the high pedestal of the strong target/clutter.
Simulation results verified the feasibility of the OMP method.

\bibliographystyle{IEEEtran}
\bibliography{Analysis_of_Random_Pulse_Repetition_Interval_Radar}

\end{document}